\documentclass[twocolumn, showpacs,amsmath,floatfix,superscriptaddress, showkeys]{revtex4-1}
\usepackage{graphicx}
\usepackage{mathptmx}
\usepackage{dcolumn}
\usepackage{bm}
\usepackage{hyperref}
\usepackage{amssymb}
\usepackage{txfonts}
\usepackage{dcolumn}
\usepackage{slashed}
\usepackage{comment}
\usepackage{mathrsfs}
\usepackage{multirow}
\usepackage{epstopdf}
\usepackage{float}
\usepackage{color}
\usepackage{indentfirst}
\usepackage[section]{placeins}
\usepackage{CJK}
\usepackage{soul}
\usepackage{ulem}

\newcommand{\delete}{\bgroup\markoverwith{\textcolor{red}{\rule[0.5ex]{2pt}{1pt}}}\ULon}

\allowdisplaybreaks

\begin{document}

%\begin{CJK}{GBK}{song}
\title{Pseudo-spin symmetry restoration and the in-medium balance between nuclear attractive and repulsive interactions}%equilibrium of nuclear dynamics}% Force line breaks with \\
%\thanks{}%

\author{Jing Geng}% (¹¢¾§)}
\affiliation{School of Nuclear Science and Technology, Lanzhou University, Lanzhou 730000, China}
\author{Jia Jie Li}% (Àî¼Ñ½Ü)}
\affiliation{School of Nuclear Science and Technology, Lanzhou University, Lanzhou 730000, China}
\affiliation{Institut f\"ur Theoretische Physik, J. W. Goethe-Universit\"at, D-60438 Frankfurt am Main, Germany}
\author{Wen Hui Long}% (ÁúÎÄ»Ô)}\email{longwh@lzu.edu.cn}
\affiliation{School of Nuclear Science and Technology, Lanzhou University, Lanzhou 730000, China}
\affiliation{Joint Department for Nuclear Physics, Lanzhou University and Institute of Modern Physics, CAS, Lanzhou 730000, China}
\affiliation{Key Laboratory of Special Function Materials and Structure Design, Ministry of Education, Lanzhou 730000, China}
\author{Yi Fei Niu}% (Å£Ò»ì³)}
\affiliation{School of Nuclear Science and Technology, Lanzhou University, Lanzhou 730000, China}
\author{Shi Yao Chang}% (³£Ê¿Ò¢)}
\affiliation{School of Nuclear Science and Technology, Lanzhou University, Lanzhou 730000, China}

\begin{abstract}
The mechanism that restores the pseudo-spin symmetry (PSS) are investigated under the relativistic Hartree-Fock (RHF) approach, by focusing on the in-medium balance between nuclear attractive and repulsive interactions. It is illustrated that the modelings of both the equilibrium of nuclear dynamics and the in-medium effects can be essentially changed by the $\rho$-tensor coupling that play the role almost fully via the Fock terms, from which the model discrepancy on the PSS restoration is verified. Specifically, the largely different density-dependent behaviors of the isoscalar coupling strengths $g_\sigma$ and $g_\omega$, deduced from the parametrization of the RHF Lagrangian PKA1, play an essential role in restoring the PSS of the high-$l'$ pseudo-spin doublets around the Fermi levels. Qualitatively, a guidance is provided for the modelings of both the equilibrium of nuclear dynamics and the in-medium effects via the PSS restoration.

\end{abstract}
 %Use showkeys class option if keyword
\maketitle

%\end{CJK}

In nuclear physics, one of the basic issues is to understand the nature of nuclear force and the consequences in nuclear structure. Among many attempts on nuclear force, the meson exchange theory proposed by Yukawa  \cite{Yukawa1935Proc.Phys.Math.Soc.Japan17.48} is still one of the most successful interpretations. From the point of view of the effective field theory, nuclear force composes of strong attractive and repulsive ingredients, which can be interpreted effectively by the exchanges of the scalar and vector mesons  \cite{Miller1972PRC5.241}. In nuclear systems, the delicate balance between the attractive and repulsive interactions leads to an equilibrium of nuclear dynamics, e.g., the residual attraction guarantees the binding of a nucleus.

Adhering to such ideal, the relativistic mean field (RMF) theory that contains only the Hartree terms of the meson exchange diagram of nuclear force, also called as the covariant density functional theory (CDFT) in recent years, has achieved many successes in describing various nuclear phenomena  \cite{Walecka1974Ann.Phys83.491, Reinhard1989Rep.Prog.Phys52.439, Ring1996Prog.Part.Nucl.Phys37.193, Bender2003Rev.Mod.Phys75.121, Vretenar2005PhyRep409.101, Meng2006Prog.Part.Nucl.Phys57.470, T.Niksic2011Prog.Part.Nucl.Phys66.519, Liang2015PRe570.1, Meng2006PRC73.037303}. As a typical evidence, the important ingredient of nuclear force --- strong spin-orbit (SO) coupling \cite{Mayer1949Phys.Rev.75.1969,Haxel1949Phys.Rev.75.1766} can be naturally interpreted, with a covariant representation of the attractive scalar potential $S(r)$ and repulsive vector one $V(r)$ of the order of several hundred MeV \cite{Meng1999PRC59.154}. Such covariant representation also works well in providing natural explanation on the origin of the pseudo-spin symmetry (PSS), the quasi-degeneracy of two single particle (s.p.) states $(n,l,j=l+1/2)$ and $(n-1,l+2,j=l+3/2)$, correspondingly the pseudo-spin (PS) doublet $(n'=n-1,l'=l+1,j'=j=l'\pm 1/2)$ \cite{Hecht1969NPA137.129,Arima1969PLB30.517}. Within the RMF scheme, the PSS is manifested as a relativistic symmetry of the Dirac Hamiltonian under the condition $S(r)+V(r) = 0$, and the pseudo-orbit $l'$ is nothing but the orbital angular momentum of the lower component of Dirac spinor \cite{Ginocchio1997PRL78.436, Ginocchio1998PRC57.1167, Ginocchio1998PLB425.1, Ginocchio1999PhyRep315.231}. Further, the condition of exact PSS was generalized as $d[S (r) + V(r)]/dr = 0$ \cite{Meng1998PRC58.R628, Meng1999PRC59.154, Chen2003CPL20.358, Zhou2003PRL91.262501}. It should be noted that both conditions imply a balance between strong nuclear attractive and repulsive interactions, which is simply evaluated by the RMF $S(r)+V(r)$. More specifically, such balance is held mainly by strong attractive $\sigma$-scalar ($\sigma$-S) and repulsive $\omega$-vector ($\omega$-V) couplings.

Under the RMF approach, the in-medium nuclear interaction, referred as the effective interaction that defines an effective Lagrangian, is able to provide accurate description of nuclear properties, by collaborating with the nonlinear self-couplings of mesons \cite{Boguta1977NPA292.413, Sugahara1994NPA579.557TM1, Long2004PRC69.034319} or the density dependencies of meson-nucleon coupling strengths \cite{Brockmann1992PRL68.3408, Lenske1995PLB345.355, Fuchs1995PRC52.3043, Typel1999NPA656.331TW99}. In fact, both are introduced to simulate the complicated in-medium effects of nuclear force. Implemented with the Fock terms \cite{Bouyssy1987PRC36.380}, an inseparable part of the meson exchange diagram of nuclear force, it was also found that the theoretical accuracy under the relativistic Hartree-Fock (RHF) approach can be largely improved by introducing the self-couplings of $\sigma$-meson or scalar fields \cite{Bernardos1993PRC48.2665, Marcos2004JPG30.703}. Moreover, assuming the meson-nucleon coupling strengths to be density-dependent, comparable accuracy as the RMF theory in describing nuclear structure has been achieved by the density-dependent relativistic Hartree-Fock (DDRHF) theory \cite{Long2006PLB640.150, Long2007PRC76.034314, Long2010PRC81.024308} with the proposed RHF Lagrangians PKO$i$ $(i = 1, 2, 3)$ \cite{Long2006PLB640.150, Long2008EPL82.12001} and PKA1 \cite{Long2007PRC76.034314}. Besides, significant improvements with the explicit treatment of the Fock terms were obtained in the self-consistent description of shell evolution \cite{Long2008EPL82.12001, Long2009PLB680.428, Li2016PLB753.97, Wang2013PRC87.047301}, nuclear tensor force \cite{Jiang2015PRC91.025802, Jiang2015PRC91.034326, Zong2018CPC42.024101}, nuclear spin-isospin excitations \cite{Liang2008PRL101.122502, Liang2009PRC79.064316, Liang2012PRC85.064302, Niu2013PLB723.172, Niu2017PRC95.044301}, symmetry energy \cite{Sun2008PRC78.065805, Long2012PRC85.025806, Zhao2015JPG42.095101}, new magicity \cite{Li2014PLB732.169, Li2016PLB753.97, Li2019PLB788.192}, etc.

Despite the successful applications, the conventional RMF models suffered much from the emergences of the spurious shell closures $N$ or $Z = 58$ and $92$, which correspond to rather large splittings of the high-$l'$ PS doublets \cite{Geng2006CPL23.1139, Long2007PRC76.034314}. For instance, the binding energies of the nuclei around $^{140}_{58}$Ce and $^{218}_{92}$U are systematically overestimated by RMF because of the existences of the spurious shells $Z=58$ and $92$ \cite{Geng2006CPL23.1139}. In fact, recent experimental measurement rules out the possibility of $Z = 92$ sub-shell \cite{M.D.Sun2017PLB771.303}. Considering the Fock terms, the missing degrees of freedom in RMF, such as the $\pi$-pseudo-vector ($\pi$-PV) and $\rho$-tensor ($\rho$-T) couplings, can be naturally included in RHF \cite{Bouyssy1987PRC36.380}. Thus, the in-medium balance between nuclear attractive and repulsive interactions is expected to be structurally changed from the RMF to RHF approaches, further impacting the PSS restoration \cite{Long2006PLB639.242}. Unfortunately, the RHF parametrizations PKO$i$ still induce such spurious shells, although both the $\sigma$-S and $\omega$-V couplings have been notably changed by the Fock terms \cite{Long2006PLB640.150}. Until implemented with the $\rho$-T coupling, these spurious shell closures are eliminated eventually by the RHF parametrization PKA1 \cite{Long2007PRC76.034314} that properly restores the PSS for the high-$l'$ PS doublets \cite{Long2009PLB680.428, Long2010PRC81.031302(R), Li2014PLB732.169}. In general for the low-$l'$ PS doublets around the Fermi levels, e.g., the $(n+1)s_{1/2}$ and $nd_{3/2}$ partners ($l'=1$), the PSS can be properly restored in the RMF calculations and the RHF ones with PKO$i$ \cite{Liang2015PRe570.1}. However for the high-$l'$ PS doublets, rather large splittings are often obtained by these calculations, in contrast to the ones with PKA1 \cite{Long2007PRC76.034314, Long2009PLB680.428, Long2010PRC81.031302(R), Li2014PLB732.169}.

\begin{figure}[htbp]
  \centering
  \includegraphics[width=0.8\linewidth]{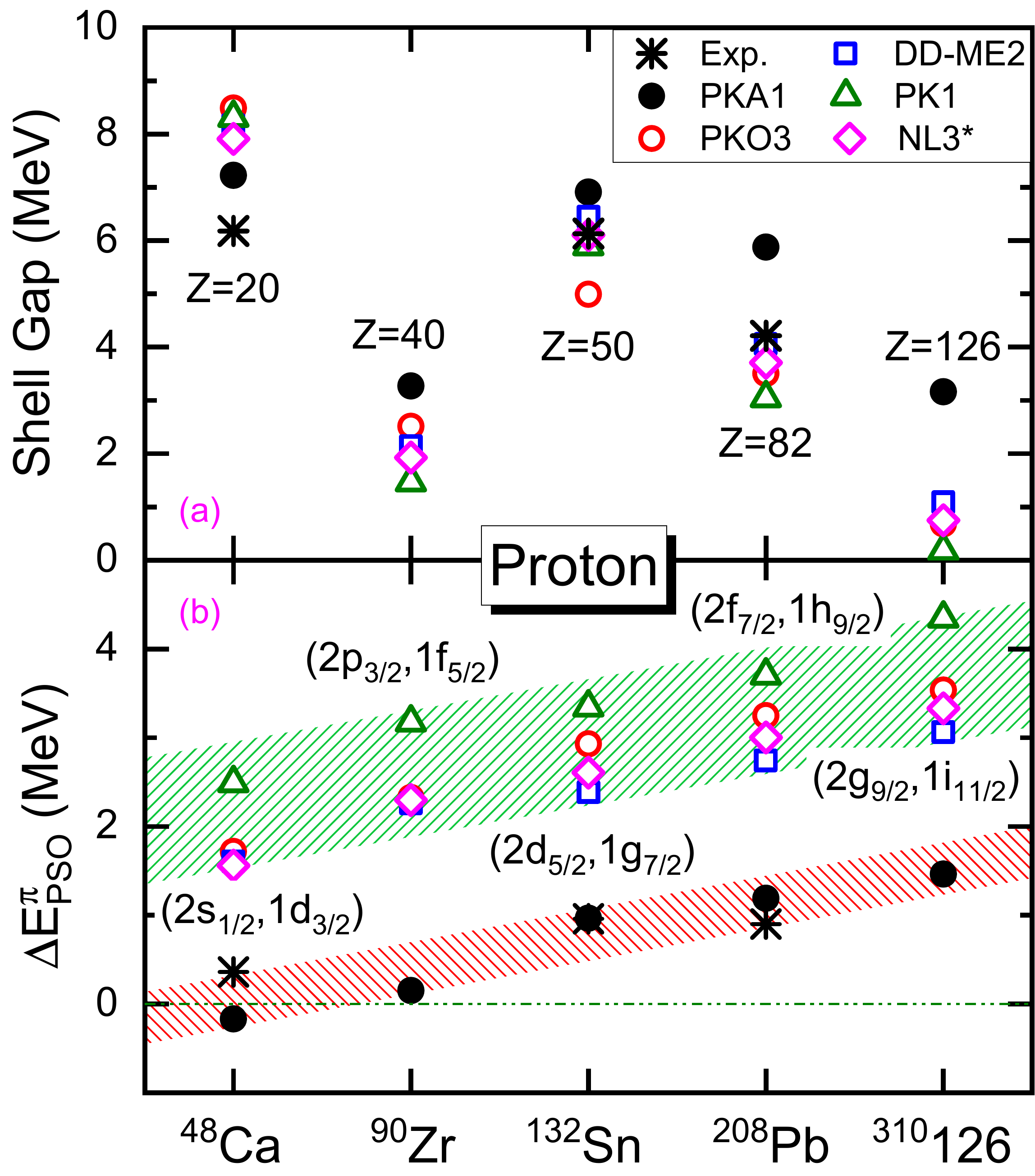}
  \caption{(Color Online) Proton shell gaps (MeV) [plot (a)] and the splittings of PS partners nearby (both above or below the Fermi levels) $\Delta E_{\text{PSO}}^{\pi}$ (MeV) [plot (b)] for the traditional magic nuclei $^{48}$Ca, $^{90}$Zr, $^{132}$Sn and $^{208}$Pb, and the superheavy one $^{310}126_{184}$ predicted by PKA1. The results are calculated by PKA1 \cite{Long2007PRC76.034314}, PKO3 \cite{Long2008EPL82.12001}, DD-ME2 \cite{Lalazissis2005PRC71.024312DDME2}, PK1 \cite{Long2004PRC69.034319} and NL3$^*$ \cite{Lalazissis2009PLB671.36NL3S}. Experimental data are taken from Ref.  \cite{Grawa2007Rep.Prog.Phys70.1525}. }\label{fig:SGandPS}
\end{figure}

Taking the traditional magic nuclei $^{48}$Ca, $^{90}$Zr, $^{132}$Sn and $^{208}$Pb, and the predicted superheavy one $^{310}$126 by PKA1 \cite{Li2014PLB732.169} as the examples, Fig. \ref{fig:SGandPS} shows the relevant proton shell gaps [plot (a)] and the splittings of the PS doublets neighboring the shells [plot (b)], given by the RHF Lagrangians PKA1 \cite{Long2007PRC76.034314} and PKO3 \cite{Long2008EPL82.12001}, and the popular RMF ones DD-ME2 \cite{Lalazissis2005PRC71.024312DDME2}, PK1 \cite{Long2004PRC69.034319} and NL3* \cite{Lalazissis2009PLB671.36NL3S}. As seen in plot (b), the pseudo-spin orbital (PSO) splittings given by all the selected Lagrangians show near parallel trends from the light to heavy nuclei. Referring to the experimental values \cite{Grawa2007Rep.Prog.Phys70.1525}, only PKA1 shows appropriate agreements, whereas the others distinctly overestimate the PSO splittings. This brings remarkable consequences on the opening of the neighboring shells in heavy nuclei.

From Fig. \ref{fig:SGandPS}, it can be seen that the proton shell $Z=82$ in $^{208}$Pb becomes less pronounced if larger splitting of the PS doublet $(2f_{7/2}, 1h_{9/2})$ is obtained by the selected Lagrangians, among which PKA1 reproduces the PSO splitting and presents the most notable magic shell \cite{Long2007PRC76.034314, Shen2013PRC88.024311}. Notice that the current calculations are performed on the level of mean field approach. If considering the effects beyond mean field approach, such as particle vibration coupling (PVC) \cite{E.V.Litvinova2011PRC84.014305}, the single-particle (s.p.) levels may be shifted towards the Fermi levels \cite{Vretenar2002PRC65.024321}. Such that the underestimated shells $Z=82$ in Fig. \ref{fig:SGandPS} (a) can be further squeezed, because the s.p. states which determine the shell gaps are either below or above the Fermi levels. On the contrary, the selected PS doublet orbits are either both above or below the Fermi levels. Qualitatively speaking, the PSO splittings in Fig. \ref{fig:SGandPS} (b) may be even enlarged slightly by the PVC effect, because it shifts the PS doublet orbits along the same direction towards the Fermi levels and such effect is reduced when the s.p. orbit is farther away from the Fermi levels \cite{Vretenar2002PRC65.024321}.

Similarly in $^{132}$Sn, the selected models, only except PKA1, also present large splittings for the PS doublet $(2d_{5/2}, 1g_{7/2})$, inducing a spurious shell $Z=58$ just above the magic one $Z=50$ \cite{Long2007PRC76.034314}. It may less impact the magicity $Z = 50$, while indeed compresses the opening of the sub-shell $Z=64$ in semi-magic nucleus $^{146}_{64}$Gd$_{82}$ \cite{Long2009PLB680.428} and limits the extension of neutron drip line of the cerium isotopes \cite{Long2010PRC81.031302(R)}. Moreover, as shown in Fig. \ref{fig:SGandPS}, the occurrence of the superheavy magicity $Z=126$ seems essentially related to the PSS restoration of the high-$l'$ PS doublet ($\pi2g_{9/2}, 1i_{11/2}$). Specifically, PKA1 predicts $Z=126$ to be magic, while the others support the magic shell $Z=138$, which corresponds to large splitting of the PS doublet ($\pi2g_{9/2}, 1i_{11/2}$). Therefore, it is worthwhile to further clarify the PSS restoration for the high-$l'$ PS doublets, in particular the distinct deviations between PKA1 and the other relativistic models.

In the following, we focus on the doubly magic nucleus $^{208}$Pb, that possesses plenty of PS doublets with various angular momenta, to study the systematics of the PSS restoration with respect to pseudo-orbit $l'$, by applying the RHF Lagrangians PKA1 and PKO3, and the RMF one DD-ME2. To verify the relation between the PSS restoration and the equilibrium of nuclear dynamics, it is necessary to briefly recall the RHF/RMF Hamiltonian, which may contain the isoscalar $\sigma$-S and $\omega$-V, and the isovector $\rho$-V, $\rho$-vector-tensor ($\rho$-VT), $\rho$-T and $\pi$-PV, and photon vector ($A$-V) couplings \cite{Bouyssy1987PRC36.380, Long2007PRC76.034314} as,
\begin{align}
\nonumber H=&\int d\boldsymbol{x} \bar{\psi} \left(-i \boldsymbol{\gamma} \cdot \boldsymbol{\nabla} + M \right) \psi\\
&\hspace{2em}+ \frac{1}{2} \sum_\phi\int d\boldsymbol{x} d\boldsymbol{x}' \bar{\psi}(\boldsymbol{x}) \bar{\psi}(\boldsymbol{x}') \Gamma_{\phi} D_{\phi} \psi(\boldsymbol{x}') \psi(\boldsymbol{x}).\label{eq:Hamiltonian}
\end{align}
In the Hamiltonian (\ref{eq:Hamiltonian}), $\psi$ represents nucleon spinors, and $D_{\phi}$ denotes the propagators in various meson (photon) coupling channels $\phi$, and the interaction vertexes $\Gamma_{\phi}$ read as,
\begin{subequations} \label{eq:vertex}
\begin{align}
\Gamma_{\sigma\text{-S}}(\boldsymbol{x},\boldsymbol{x}')\equiv&-\left(g_{\sigma}\right)_{\boldsymbol{x}}\left(g_{\sigma}\right)_{\boldsymbol{x}'};\\
\Gamma_{\omega\text{-V}}(\boldsymbol{x},\boldsymbol{x}')\equiv&\left(g_{\omega}\gamma_{\mu}\right)_{\boldsymbol{x}}\left(g_{\omega}\gamma^{\mu}\right)_{\boldsymbol{x}'};\\
\Gamma_{\rho\text{-V}}(\boldsymbol{x},\boldsymbol{x}')\equiv&\left(g_{\rho}\gamma_{\mu}\vec{\tau}\right)_{\boldsymbol{x}}\cdot \left(g_{\rho}\gamma^{\mu}\vec{\tau}\right)_{\boldsymbol{x}'};\\
\Gamma_{\text{A}\text{-V}}(\boldsymbol{x},\boldsymbol{x}')\equiv&\frac{e^{2}}{4}\left[\gamma_{\mu}\left(1-\tau_{3}\right)\right]_{\boldsymbol{x}} \left[\gamma^{\mu}\left(1-\tau_{3}\right)\right]_{\boldsymbol{x}'};\\
\Gamma_{\pi\text{-PV}}(\boldsymbol{x},\boldsymbol{x}')\equiv&\frac{-1}{m_{\pi}^{2}}\left(f_{\pi}\vec{\tau}\gamma_{5}\gamma_{k}\partial^{k}\right)_{\boldsymbol{x}}\cdot \left(f_{\pi} \vec{\tau}\gamma_{5}\gamma_{l}\partial^{l}\right)_{\boldsymbol{x}'};\\
\Gamma_{\rho\text{-T}}(\boldsymbol{x},\boldsymbol{x}')\equiv&\frac{1}{4M^{2}}\left(f_{\rho}\sigma_{\nu k}\vec{\tau}\partial^{k}\right)_{\boldsymbol{x}}\cdot\left(f_{\rho}\sigma^{\nu l}\vec{\tau}\partial_{l}\right)_{\boldsymbol{x}'};\\
\nonumber\Gamma_{\rho\text{-VT}}(\boldsymbol{x},\boldsymbol{x}')\equiv&\frac{1}{2M}\left(f_{\rho}\sigma^{k\nu}\vec{\tau}\partial_{k}\right)_{\boldsymbol{x}}\cdot\left(g_{\rho}\gamma_{\nu}\vec{\tau}\right)_{\boldsymbol{x}'}\\
&+\left(g_{\rho}\gamma_{\nu}\vec{\tau}\right)_{\boldsymbol{x}}\cdot\frac{1}{2M}\left(f_{\rho}\sigma^{k\nu}\vec{\tau}\partial_{k}\right)_{\boldsymbol{x}'}.
\end{align}
\end{subequations}
Here the bold type represents the vectors in space and arrows for the isovector ones. In deriving the Hamiltonian (\ref{eq:Hamiltonian}), the retardation effects are neglected in the propagators $D_\phi$ as usual. To get the energy functional, namely the expectation of the Hamiltonian with respect to the Hartree-Fock ground state \cite{Bouyssy1987PRC36.380}, Dirac spinor $\psi$ is quantized in terms of the creation and annihilation operators defined by the solutions of the Dirac equation. Restricted on the level of mean field approach, the contributions from the negative energy states in quantizing the Dirac spinor $\psi$ are ignored, namely the no-sea approximation. Thus, the energy functional can be obtained from the expectation of the Hamiltonian (\ref{eq:Hamiltonian}) with respect to the Hartree-Fock ground state \cite{Bouyssy1987PRC36.380}, in which the Fock terms are considered/dropped explicitly in the RHF/RMF approach.

In order to provide accurate description of nuclear properties, the coupling strengths $g_{\sigma}$, $g_{\omega}$, $g_{\rho}$, $f_{\pi}$ and $f_{\rho}$ in the vertexes (\ref{eq:vertex}) are treated as functions of the nucleon density $\rho_{b}$ to take the nuclear in-medium effects into account phenomenologically. For the details, please refer to Refs. \cite{Brockmann1992PRL68.3408, Lenske1995PLB345.355, Typel1999NPA656.331TW99, Long2006PLB640.150, Long2007PRC76.034314}.

\begin{table}[h]
\caption{Contributions to the binding energy (MeV) of $^{208}$Pb from various channels given by the RHF Lagrangian PKA1, where $E_{\text{kin.}}$ and $E_{\text{c.m.}}$ correspond to the kinetic energy and the center-of-mass corrections, respectively.} \label{tab:ENE}%\renewcommand{\arraystretch}{1.2}
\begin{tabular}{c|c|rrr}\hline\hline
     \multicolumn{2}{c|}{$^{208}$Pb}    &  Neutron~~& Proton~~~  &  Total~~~   \\ \hline
  \multicolumn{2}{c|}{$E_{\text{kin.}}$}&   1596.56 &     907.90 &    2504.45  \\ \hline
\multirow{6}{*}{Hartree}    &$\sigma$-S & -14321.01 &  -10022.61 &  -24343.62  \\
                            & $\omega$-V&  11520.95 &    7962.61 &   19483.56  \\
                            & $A$-V     &      0.00 &     827.64 &     827.64  \\ \cline{2-5}
                            & $\rho$-V  &     98.42 &     -65.12 &      33.30  \\
                            & $\rho$-VT &     -1.65 &       1.08 &      -0.57  \\
                            & $\rho$-T  &     -0.31 &       0.21 &      -0.10  \\ \hline\hline
\multirow{7}{*}{ Fock}      & $\sigma$-S&   3503.28 &    1774.31 &    5277.58  \\
                            & $\omega$-V&  -2451.24 &   -1265.22 &   -3716.45  \\
                            & $A$-V     &      0.00 &     -29.02 &     -29.02  \\ \cline{2-5}
                            & $\rho$-V  &   -266.20 &    -210.10 &    -476.30  \\
                            & $\rho$-VT &    122.51 &      89.16 &     211.66  \\
                            & $\rho$-T  &   -687.90 &    -531.94 &   -1219.85  \\%\cline{2-5}
                            & $\pi$     &   -103.26 &     -79.75 &    -183.01  \\ \hline
  \multicolumn{2}{c|}{$E_{\text{c.m.}}$}&     -3.34 &      -2.24 &      -5.58  \\ \hline
  \multicolumn{2}{c|}{Total}            &   -993.17 &    -643.11 &   -1636.28  \\
\hline\hline
\end{tabular}

\end{table}

Under the RMF approach, the $\sigma$-S and $\omega$-V couplings represent respectively the strong attractive and repulsive nuclear interactions, and the $\rho$-V coupling, as well as the $\delta$-scalar ($\delta$-S) one \cite{Roca-Maza2011PRC84.054309}, characterizes the isovector nature, and the $A$-V coupling for the Coulomb interaction. Further considering the Fock terms explicitly, namely the RHF approach, the RHF Lagrangians PKO1 and PKO3 take the $\pi$-PV coupling into account, and PKA1 contains both the $\pi$-PV and $\rho$-T coupling (automatically the $\rho$-VT one) \cite{Long2006PLB640.150, Long2007PRC76.034314, Long2008EPL82.12001}. To have a complete understanding on the energy functional, Table \ref{tab:ENE} shows the kinetic energy $E_{\text{kin.}}$, the potential energies of various channels and the center-of-mass corrections $E_{\text{c.m.}}$ \cite{Long2004PRC69.034319} given by PKA1 for $^{208}$Pb.

Generally speaking, the equilibrium of nuclear dynamics, e.g., the binding of nucleus, is dominated by the strong isoscalar $\sigma$-S and $\omega$-V couplings. For the isovector channels, the Hartree terms (see the upper panel of Table \ref{tab:ENE}) present tiny contributions after the cancellation between neutron and proton. Thus, the balance between the $\sigma$-S and $\omega$-V couplings dominates the equilibrium of nuclear dynamics in the RMF approach. Because of the Fock terms, the isovector channels present net attractive and notably enlarged mean field potentials. In particular, the ones contributed by the $\rho$-T coupling are rather strongly attractive, even after cancelling with the repulsive $\rho$-VT coupling. From Table \ref{tab:ENE}, it is obvious that the $\pi$-PV and $\rho$-T ($\rho$-VT) couplings play the role almost fully via the Fock terms, and the $\sigma$-S and $\omega$-V couplings are the dominant channels in deducing the equilibrium of nuclear dynamics, although the isovector Fock terms present substantial contributions.

\begin{figure}[h]
  \centering
  \includegraphics[width=0.9\linewidth]{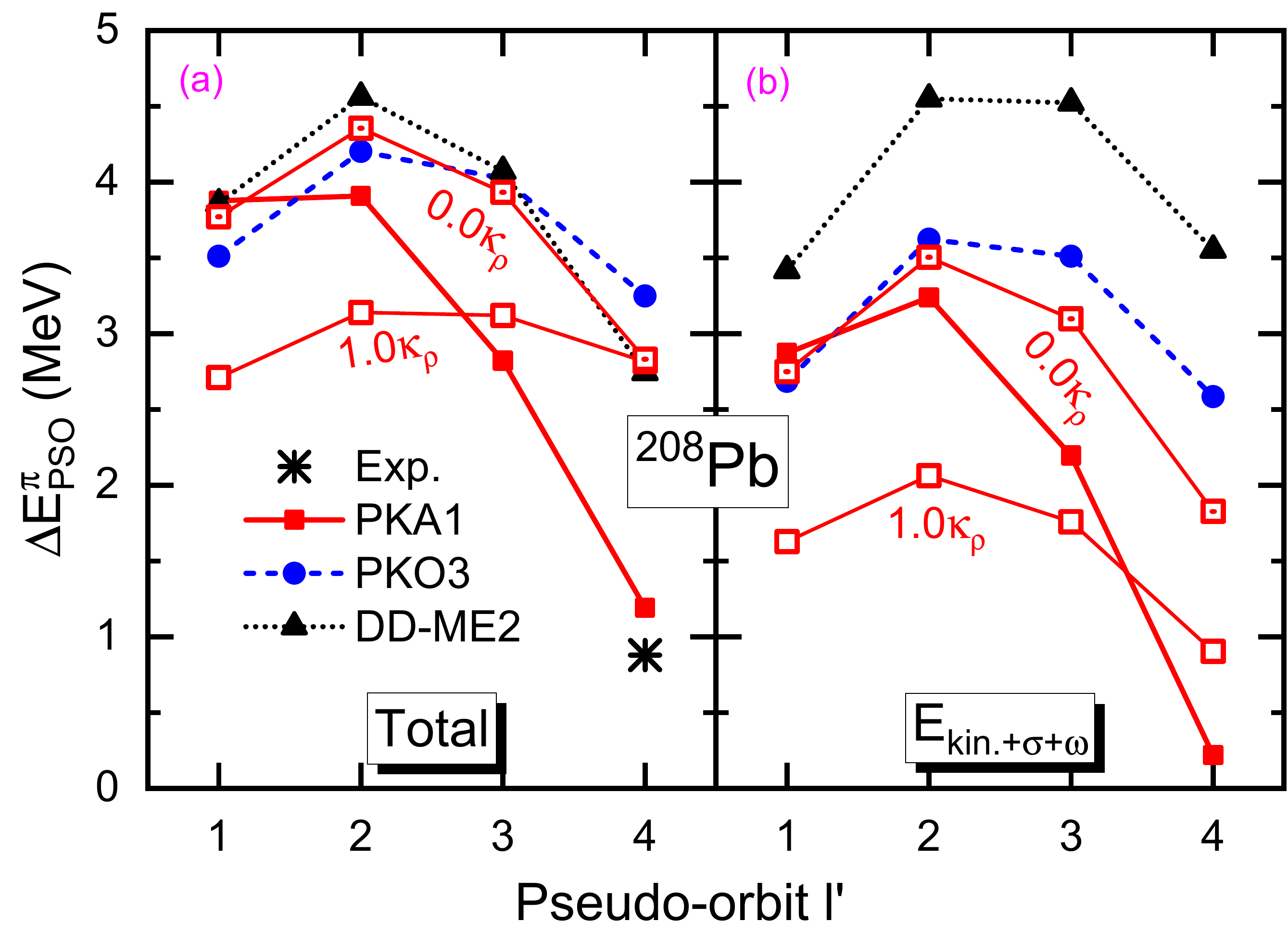}
  \caption{(Color Online) Proton ($\pi$) pseudo-spin orbital (PSO) splittings $\Delta E_{\text{PSO}}^{\pi}$ (MeV) in $^{208}$Pb as functions of pseudo-orbit $l'$ [plot (a)], as well as the sum contributions from kinetic energy, $\sigma$ and $\omega$ potential energies $E_{\text{kin.}+\sigma+\omega}$ [plot (b)]. The results are extracted from the calculations with PKA1, PKO3, DD-ME2, and the tentative parametrizations 1.0$\kappa_\rho$ and 0.0$\kappa_\rho$ based on PKA1. }\label{fig:splitting}
\end{figure}

Using PKA1, PKO3 and DD-ME2, Fig. \ref{fig:splitting} shows the proton $(\pi)$ PSO splittings $\Delta E_{\text{PSO}}^{\pi}$ in $^{208}$Pb as functions of pseudo-orbit $l'$, including the PS doublets $(\pi2s_{1/2},\pi1d_{3/2})$, $(\pi2p_{3/2},\pi1f_{5/2})$, $(\pi2d_{5/2},\pi1g_{7/2})$ and $(\pi2f_{7/2},\pi1h_{9/2})$ with $l'$ from $1$ to $4$, respectively. The experimental value for $l' = 4$ \cite{Grawa2007Rep.Prog.Phys70.1525} is also shown (snowflake symbol) in plot (a) for reference. As revealed in Refs.  \cite{Meng1998PRC58.R628,Meng1999PRC59.154,Long2006PLB639.242}, the PSO splittings decrease towards the Fermi levels in general, leading to properly restored PSS around Fermi levels. Coincidentally, the $\Delta E_{\text{PSO}}^{\pi}$ given by PKA1 decreases with respect to $l'$ towards the Fermi level, and consistent with the experimental value a well restored PSS at $l' = 4$ is achieved by PKA1. On the contrary, both PKO3 and DD-ME2 present rather large $\Delta E_{\text{PSO}}^{\pi}$ values at $l' = 4$, corresponding to the spurious shell $Z = 92$ \cite{Geng2006CPL23.1139,M.D.Sun2017PLB771.303}.

We further show in Fig. \ref{fig:splitting} (b) the sum contributions to $\Delta E_{\text{PSO}}^{\pi}$ from the kinetic term, and the dominant $\sigma$-S and $\omega$-V channels, denoted as $E_{\text{kin.}+\sigma+\omega}$. As mentioned, the equilibrium of nuclear dynamics is determined mainly by the balance between the strong attractive $\sigma$-S and repulsive $\omega$-V couplings. In Fig. \ref{fig:splitting} (b), it is also clearly shown that the systematics of $\Delta E_{\text{PSO}}^{\pi}$ can be interpreted almost fully by the sum contributions $E_{\text{kin.}+\sigma+\omega}$. Consistent with Fig. \ref{fig:splitting} (a), the $E_{\text{kin.}+\sigma+\omega}$ term given by PKA1 decrease distinctly with respect to $l'$, in contrast to PKO3 and DD-ME2. One may also notice that from the RMF (DD-ME2) to the RHF (PKO3 and PKA1) models, the $E_{\text{kin.}+\sigma+\omega}$ contributions are remarkably reduced. Specifically, the $l'$-dependencies are shifted nearly in parallel from DD-ME2 to PKO3. Both may imply some systematical changes from the RMF to RHF approaches in modeling the equilibrium of nuclear dynamics.

\begin{figure}[htbp] \centering
  \includegraphics[width=0.9\linewidth]{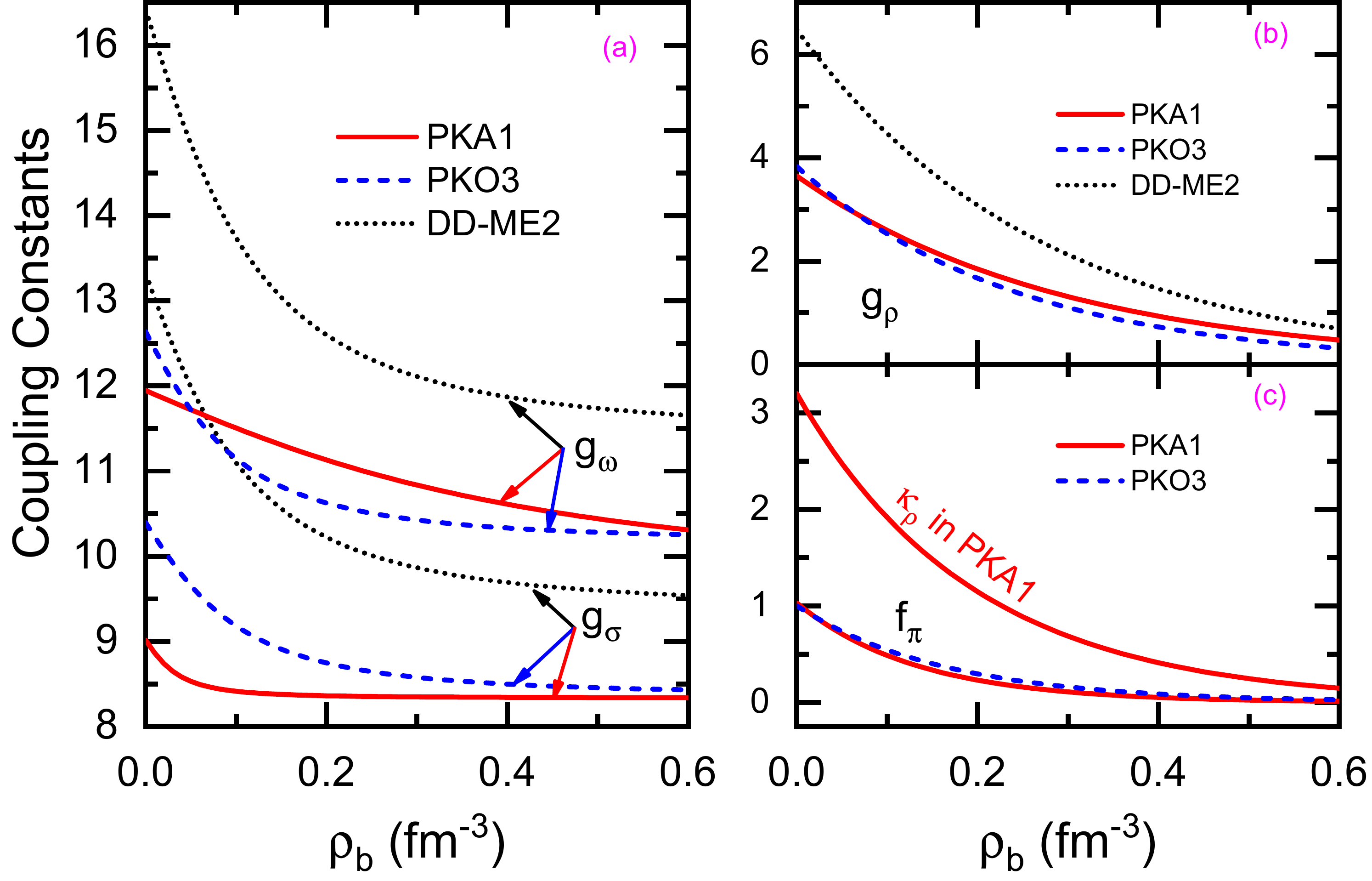}\\
  \caption{(Color Online) Meson-nucleon coupling constants, namely the isoscalar $g_\sigma$ and $g_\omega$ [plot (a)], the isovector $g_\rho$ [plot (b)], $\kappa_\rho$ [$\kappa_\rho(0) = f_\rho(0)/g_\rho(0)$] and $f_\pi$ [plot (c)],  as functions of the density $\rho_b$ (fm$^{-3}$) for PKA1, PKO3 and DD-ME2.}\label{fig:gsw}
\end{figure}

According to the conditions $V(r)+S(r)=0$ and $d[V(r)+S(r)]/dr =0 $, the PSS restoration is governed by the delicate cancellation between nuclear attractive and repulsive interactions, as well as the equilibrium of nuclear dynamics. Thus, the model discrepancy shown in Fig. \ref{fig:splitting} can be traced back to the details of the in-medium nuclear interactions. In this context, the modeling of nuclear in-medium effects, i.e., by the density dependence of the coupling strengths, may also result noticeable effects. To further understand the PSS restoration for the high-$l'$ PS doublets, we show in Table \ref{tab:sumene} the energy functional of $^{208}$Pb, including the total binding energy $E_{\text{Total}}$, and the contributions from the term $E_{\text{kin.}+\sigma+\omega}$, the isovector channels ($\rho$- and $\pi$-couplings) $E_{\rho+\pi}$ and Coulomb field $E_{\text{cou.}}$. Moreover, Fig. \ref{fig:gsw} shows the isoscalar coupling strengths $g_\sigma$ and $g_\omega$ (left panel), and the isovector ones $g_\rho$, $\kappa_\rho$ [$\kappa_\rho(0)=f_\rho(0)/g_\rho(0)$] and $f_\pi$ (right panels) as functions of the density $\rho_b$.

\begin{table}[htbp]
\caption{Contributions to the binding energy (MeV) of $^{208}$Pb from the kinetic and isoscalar potential energies ($E_{\text{kin.}+\sigma+\omega}$), the isovector potential energies ($E_{\rho+\pi}$) and the Coulomb ones ($E_{\text{cou.}}$), calculated by PKA1, PKO3, DD-ME2, and the tentative parametrizations 1.0$\kappa_\rho$ and 0.0$\kappa_\rho$ based on PKA1. } \label{tab:sumene}\setlength{\tabcolsep}{0.5em}%\renewcommand{\arraystretch}{1.2}
\begin{tabular}{c|rrr|r}  \hline\hline

                      & $E_{\text{kin.}+\sigma+\omega}$  &$E_{\rho+\pi}$ & $E_{\text{cou.}}$  & $E_{\text{Total}}$ \\ \hline
 DD-ME2               &    $-$2559.81                    &        100.27 &            827.23  &  $-$1637.39      \\
 PKO3                 &    $-$1781.19                    &     $-$648.75 &            798.38  &  $-$1636.80      \\%
 PKA1                 &     $-$795.09                    &    $-$1634.84 &            798.62  &  $-$1636.27        \\
 \hline\hline
 1.0$\kappa_{\rho}$   &     $-$539.92                    &    $-$1826.54 &            734.61  &  $-$1636.46      \\
 0.0$\kappa_{\rho}$   &    $-$1764.32                    &     $-$640.81 &            773.45  &  $-$1636.66      \\
\hline\hline
\end{tabular}
\end{table}

In general, various RHF/RMF Lagrangians can give similar total binding energy for doubly magic $^{208}$Pb, a reference nucleus to calibrate the model, see the last column of Table \ref{tab:sumene}. The Coulomb potential energies are also quite similar, and the slight differences between the RMF DD-ME2 and RHF ones are due to the Fock terms. However, specific contributions from the meson coupling channels are quite different. For the RMF Lagrangian DD-ME2, the binding energy is dominated by the term $E_{\text{kin}.+\sigma+\omega}$, in which there exists large cancellation between the dominant $\sigma$-S and $\omega$-V channels as seen in Table \ref{tab:ENE}. From DD-ME2 to the RHF one PKO3, the isovector channels present a net attractive contributions due to the Fock terms, showing a notably enlarged $E_{\rho+\pi}$ value. Consistently, the term $E_{\text{kin}.+\sigma+\omega}$ is largely reduced to maintain the equilibrium of nuclear dynamics, here giving similar binding energy of $^{208}$Pb. Even though, the term $E_{\text{kin}.+\sigma+\omega}$ still remains dominant in determining the binding of $^{208}$Pb for DD-ME2 and PKO3. %Such model alteration can be easily understood from the fact that the isovector Fock terms are attractive, whereas the Hartree terms, e.g., the ones in DD-ME2, present counteracting contributions for neutron and proton, see Table \ref{tab:ENE}.
However, because the $\rho$-T coupling presents rather strong attractive potential, the isovector energy functional $E_{\rho+\pi}$ in PKA1 is much more enhanced with an even large value than the sum $E_{\text{kin}.+\sigma+\omega}$, see Table \ref{tab:sumene}. This indicates that the balance between the dominant $\sigma$-S and $\omega$-V channels is significantly modified by the $\rho$-T coupling in PKA1.

Coincident with the model alterations in determining the equilibrium of nuclear dynamics, essential changes from the RMF to RHF approaches are also found on modeling the nuclear in-medium effects, here by the density dependencies of the coupling strengths shown in Fig. \ref{fig:gsw}. As seen from Tables \ref{tab:ENE} and \ref{tab:sumene}, the isovector potential energies are enhanced distinctly by the Fock terms. One may expect that more nuclear in-medium effects can be enclosed in the isovector channels. Thus, to maintain the appropriate modelings of both the equilibrium of nuclear dynamics and the whole in-medium effects, both the values and density dependencies of the dominant coupling strengths $g_\sigma$ and $g_\omega$ are reduced from DD-ME2 to PKO3, as seen from Fig. \ref{fig:gsw} (a). Similar reduction on $g_\rho$ is also found in Fig. \ref{fig:gsw} (b). Meanwhile, it is also interestingly seen that the density-dependent behaviors of $g_{\sigma}$ and $g_{\omega}$ can be shifted almost in parallel between each another for both PKO3 and DD-ME2, namely $g_\sigma(\rho_b)/ g_\omega(\rho_b)\approx C_0$ ($C_0$ being a density independent constant). In fact, such approximate relation can be commonly found in the other popular density-dependent RMF Lagrangians and the RHF PKO$i$.

However for PKA1, due to rather strong attractive potential contributed by the $\rho$-T coupling (see Tables \ref{tab:ENE} and \ref{tab:sumene}), it indeed shakes the balance between the $\sigma$-S and $\omega$-V couplings. As shown in Fig. \ref{fig:gsw} (c), the coupling strength $\kappa_\rho$ [$\kappa_\rho(0) = f_\rho(0)/g_\rho(0)$] is also strongly density-dependent, and more nuclear in-medium effects are then carried by the $\rho$-couplings ($\rho$-V, $\rho$-T and $\rho$-VT) in PKA1 than that in PKO3, since both share similar $g_\rho$ and $f_\pi$ as seen from plots (b) and (c). Consistently, to maintain the modelings of the equilibrium of nuclear dynamics and the in-medium effects, the density dependencies of $g_\sigma$ and $g_\omega$, as well as the average value of $g_\sigma$, are further reduced from PKO3 to PKA1. In particular, as seen from Fig. \ref{fig:gsw} (a), the density-dependent behaviors of $g_{\sigma}$ and $g_{\omega}$ in PKA1 are rather different from each another, in contrast to the approximate relation $g_\sigma(\rho_b)/g_\omega(\rho_b)\approx C_0$ deduced by the parametrizations DD-ME2 and PKO3. Following the model alterations, one can conclude that the modelings of both the equilibrium of nuclear dynamics and the in-medium effects are structurally changed from the RMF to RHF approaches, in which the Fock terms, especially the $\rho$-T coupling, play an essential role.

Notice that these model deviations are innately rooted from the parameterizations of the RMF/RHF Lagrangians. This inspires us to perform some tentative parametrizations to clarify the model discrepancy on the PSS restoration shown in Figs. \ref{fig:SGandPS} and \ref{fig:splitting}. Since PKA1 presents very different description on the PSS restoration of the high-$l'$ PS doublets from DD-ME2 and PKO3, we choose PKA1 as the starting point of the parametrizations. Because the density-dependent behaviors of $g_\sigma$ and $g_\omega$ in PKA1 are quite different, as an approximation to DD-ME2 and PKO3, we firstly choose same density dependence for $g_\sigma$ and $g_\omega$. Specifically, starting from PKA1, the density dependence of $g_{\omega}$ is replaced by that of $g_{\sigma}$, namely $g_\omega(\rho_b) = g_\omega(\rho_0)/g_\sigma(\rho_0) g_\sigma(\rho_b)$ ($\rho_{0}$ being the saturation density). To approximately preserve the equilibrium of nuclear dynamics, $g_{\sigma}(\rho_{0})$ and $g_{\omega}(\rho_{0})$ are simultaneously adjusted by few percents $(\sim 4\%)$ to give identical total binding energy of $^{208}$Pb and the obtained parameter set is denoted as $1.0\kappa_{\rho}$. Subsequently, the $\rho$-T coupling is switched off, and $g_{\sigma}(\rho_{0})$ and $g_{\omega}(\rho_{0})$ are further adjusted simultaneously to maintain the binding energy of $^{208}$Pb unchanged, leading to the set $0.0\kappa_{\rho}$. The corresponding results are shown in Fig. \ref{fig:splitting} (open symbols) and the last two rows of Table \ref{tab:sumene}.

As shown in Fig. \ref{fig:splitting}, the PSO splittings $\Delta E_{\text{PSO}}^{\pi}$ are largely changed from PKA1 to the sets $1.0\kappa_{\rho}$ and $0.0\kappa_{\rho}$, and eventually similar $\Delta E_{\text{PSO}}^{\pi}$ values are obtained by the set $0.0\kappa_\rho$ as DD-ME2 and PKO3. Specifically, with identical density dependence of $g_{\sigma}$ and $g_{\omega}$, both sets $1.0\kappa_{\rho}$ and $0.0\kappa_{\rho}$ present much weaker $l'$-dependence than the original PKA1 on both total $\Delta E_{\text{PSO}}^{\pi}$ and the dominant $E_{\text{kin.}+\sigma+\omega}$ contributions. Coincidentally for both PKO3 and DD-ME2, which have $g_\sigma(\rho_b)/g_\omega(\rho_b)\approx C_0$, also present weak $l'$-dependent $E_{\text{kin.}+\sigma+\omega}$ terms. It seems that the $l'$-dependence of $\Delta E_{\text{PSO}}^{\pi}$ is sensitive to the difference between the density-dependent behaviors of $g_{\sigma}$ and $g_{\omega}$, i.e., more difference leads to stronger $l'$-dependence. This is not quite difficult to understand from the in-medium effects. In realistic nuclei, the centrifugal repulsion is enhanced with angular momentum increasing, and nucleons are driven farther away from the center to the surface of nucleus, where nucleon density varies from roughly saturated to zero values. Due to largely different density-dependent behaviors of $g_\sigma$ and $g_\omega$, the balance between $\sigma$-S and $\omega$-V couplings described by PKA1 is also sensitively changed from the center to the surface regions, corresponding to the residual in-medium effects in the isoscalar channels, and such effects are manifested as strongly $l'$-dependent $\Delta E_{\text{PSO}}^\pi$ values in Fig. \ref{fig:splitting}

Moreover, as shown in Table \ref{tab:sumene}, the $E_{\text{kin.}+\sigma+\omega}$ values are remarkably enlarged from the set $1.0\kappa_{\rho}$ to $0.0\kappa_{\rho}$, and accordingly the contributions to $\Delta E_{\text{PSO}}^\pi$ are raised systematically as seen from Fig. \ref{fig:splitting} (b). In fact, such alteration can be interpreted qualitatively by the condition of the exact PSS, namely $S(r) + V(r) = 0$ or $d[S(r)+V(r)]/dr=0$, that more cancellation between the attractive and repulsive potentials leads to smaller PSO splitting, and vice versa. Similarly, smaller $E_{\text{kin.}+\sigma+\omega}$ value is obtained by PKO3 than DD-ME2 as shown in Table \ref{tab:sumene} and reduced $E_{\text{kin.}+\sigma+\omega}$ contributions to $\Delta E_{\text{PSO}}^\pi$ are found in Fig. \ref{fig:splitting} (b) from DD-ME2 to PKO3.

At the end, it shall be mentioned that the parametrizations 1.0$\kappa_\rho$ and 0.0$\kappa_\rho$ are not fully performed and the consequent effects to other observables, such as the energies or the radii of other nuclei in nuclear chart, are not discussed. Combined with the systematical alterations from PKA1 to PKO3 and further to DD-ME2, such full parametrizations are not necessary indeed. One may also notice that the set $0.0\kappa_\rho$, in which the $\rho$-T coupling is switched off, shares same degrees of freedom as PKO3. In Table \ref{tab:sumene}, the set $0.0\kappa_{\rho}$ presents roughly identical $E_{\text{kin.}+\sigma+\omega}$ and $E_{\rho+\pi}$ values as PKO3, as well as similar $E_{\text{kin.}+\sigma+\omega}$ contributions to $\Delta E_{\text{PSO}}^\pi$ in Fig. \ref{fig:splitting} (b). Thus, similar as the trend from PKA1 to PKO3, the conclusion indicated by the systematics from PKA1 to the set $0.0\kappa_\rho$ will not be much changed, even performing the full parametrization for the set $0.0\kappa_\rho$, because it can converge to PKO3 if applying similar parametrization conditions as PKO3, and so it does from PKO3 to DD-ME2, if further excluding the Fock terms.

In summary, under the relativistic Hartree-Fock (RHF) approach, we studied the in-medium balance between nuclear attractive and repulsive interactions, and the consequences on the restoration of the pseudo-spin symmetry (PSS). As compared to the traditional relativistic mean field models, the in-medium balance between attractive $\sigma$-S and repulsive $\omega$-V couplings is largely changed by the Fock terms, especially by the $\rho$-T coupling in the RHF Lagrangian PKA1. Coincidentally on the modeling of nuclear in-medium effects, the parametrization PKA1 leads to rather different density-dependent behaviors of the coupling strengths $g_{\sigma}$ and $g_{\omega}$. Both are crucial for the appropriate PSS restoration of the high-$l'$ pseudo-spin (PS) doublets around the Fermi levels, referring to the experimental data. On the level of mean field approach, our results provide a qualitative guidance on the modelings of both the equilibrium of nuclear dynamics and the in-medium effects from the aspect of the PSS restoration.

This work is partly supported by the National Natural Science Foundation of China under Grant Nos. 11675065, 11875152 and 11711540016, and the authors also want to thank Prof. Peter Ring for fruitful discussions.
%%
%\bibliographystyle{apsrev}
%\bibliography{Reference}

\begin{thebibliography}{64}
\expandafter\ifx\csname natexlab\endcsname\relax\def\natexlab#1{#1}\fi
\expandafter\ifx\csname bibnamefont\endcsname\relax
  \def\bibnamefont#1{#1}\fi
\expandafter\ifx\csname bibfnamefont\endcsname\relax
  \def\bibfnamefont#1{#1}\fi
\expandafter\ifx\csname citenamefont\endcsname\relax
  \def\citenamefont#1{#1}\fi
\expandafter\ifx\csname url\endcsname\relax
  \def\url#1{\texttt{#1}}\fi
\expandafter\ifx\csname urlprefix\endcsname\relax\def\urlprefix{URL }\fi
\providecommand{\bibinfo}[2]{#2}
\providecommand{\eprint}[2][]{\url{#2}}

\bibitem[{\citenamefont{Yukawa}(1935)}]{Yukawa1935Proc.Phys.Math.Soc.Japan17.48}
\bibinfo{author}{\bibfnamefont{H.}~\bibnamefont{Yukawa}},
  \bibinfo{journal}{Proc. Phys. Math. Soc. Japan}
  \textbf{\bibinfo{volume}{17}}, \bibinfo{pages}{48} (\bibinfo{year}{1935}).

\bibitem[{\citenamefont{Miller and Green}(1972)}]{Miller1972PRC5.241}
\bibinfo{author}{\bibfnamefont{L.~D.} \bibnamefont{Miller}} \bibnamefont{and}
  \bibinfo{author}{\bibfnamefont{A.~E.~S.} \bibnamefont{Green}},
  \bibinfo{journal}{Phys. Rev. C} \textbf{\bibinfo{volume}{5}},
  \bibinfo{pages}{241} (\bibinfo{year}{1972}).

\bibitem[{\citenamefont{Walecka}(1974)}]{Walecka1974Ann.Phys83.491}
\bibinfo{author}{\bibfnamefont{J.~D.} \bibnamefont{Walecka}},
  \bibinfo{journal}{Ann. Phys. (NY)} \textbf{\bibinfo{volume}{83}},
  \bibinfo{pages}{491} (\bibinfo{year}{1974}).

\bibitem[{\citenamefont{Reinhard}(1989)}]{Reinhard1989Rep.Prog.Phys52.439}
\bibinfo{author}{\bibfnamefont{P.-G.} \bibnamefont{Reinhard}},
  \bibinfo{journal}{Rep. Prog. Phys} \textbf{\bibinfo{volume}{52}},
  \bibinfo{pages}{439} (\bibinfo{year}{1989}).

\bibitem[{\citenamefont{Ring}(1996)}]{Ring1996Prog.Part.Nucl.Phys37.193}
\bibinfo{author}{\bibfnamefont{P.}~\bibnamefont{Ring}}, \bibinfo{journal}{Prog.
  Part. Nucl. Phys} \textbf{\bibinfo{volume}{37}}, \bibinfo{pages}{193}
  (\bibinfo{year}{1996}).

\bibitem[{\citenamefont{Bender et~al.}(2003)\citenamefont{Bender, Heenen, and
  Reinhard}}]{Bender2003Rev.Mod.Phys75.121}
\bibinfo{author}{\bibfnamefont{M.}~\bibnamefont{Bender}},
  \bibinfo{author}{\bibfnamefont{P.-H.} \bibnamefont{Heenen}},
  \bibnamefont{and} \bibinfo{author}{\bibfnamefont{P.-G.}
  \bibnamefont{Reinhard}}, \bibinfo{journal}{Rev. Mod. Phys}
  \textbf{\bibinfo{volume}{75}}, \bibinfo{pages}{121} (\bibinfo{year}{2003}).

\bibitem[{\citenamefont{Vretenar et~al.}(2005)\citenamefont{Vretenar,
  Afanasjev, Lalazissis, and Ring}}]{Vretenar2005PhyRep409.101}
\bibinfo{author}{\bibfnamefont{D.}~\bibnamefont{Vretenar}},
  \bibinfo{author}{\bibfnamefont{A.~V.} \bibnamefont{Afanasjev}},
  \bibinfo{author}{\bibfnamefont{G.~A.} \bibnamefont{Lalazissis}},
  \bibnamefont{and} \bibinfo{author}{\bibfnamefont{P.}~\bibnamefont{Ring}},
  \bibinfo{journal}{Phys. Rep} \textbf{\bibinfo{volume}{409}},
  \bibinfo{pages}{101} (\bibinfo{year}{2005}).

\bibitem[{\citenamefont{Meng et~al.}(2006{\natexlab{a}})\citenamefont{Meng,
  Toki, Zhou, Zhang, Long, and Geng}}]{Meng2006Prog.Part.Nucl.Phys57.470}
\bibinfo{author}{\bibfnamefont{J.}~\bibnamefont{Meng}},
  \bibinfo{author}{\bibfnamefont{H.}~\bibnamefont{Toki}},
  \bibinfo{author}{\bibfnamefont{S.~G.} \bibnamefont{Zhou}},
  \bibinfo{author}{\bibfnamefont{S.~Q.} \bibnamefont{Zhang}},
  \bibinfo{author}{\bibfnamefont{W.~H.} \bibnamefont{Long}}, \bibnamefont{and}
  \bibinfo{author}{\bibfnamefont{L.~S.} \bibnamefont{Geng}},
  \bibinfo{journal}{Prog. Part. Nucl. Phys} \textbf{\bibinfo{volume}{57}},
  \bibinfo{pages}{470} (\bibinfo{year}{2006}{\natexlab{a}}).

\bibitem[{\citenamefont{Nik$\check{\text{s}}$i$\acute{\text{c}}$
  et~al.}(2011)\citenamefont{Nik$\check{\text{s}}$i$\acute{\text{c}}$,
  Vretenar, and Ring}}]{T.Niksic2011Prog.Part.Nucl.Phys66.519}
\bibinfo{author}{\bibfnamefont{T.}~\bibnamefont{Nik$\check{\text{s}}$i$\acute{\text{c}}$}},
  \bibinfo{author}{\bibfnamefont{D.}~\bibnamefont{Vretenar}}, \bibnamefont{and}
  \bibinfo{author}{\bibfnamefont{P.}~\bibnamefont{Ring}},
  \bibinfo{journal}{Prog. Part. Nucl. Phys} \textbf{\bibinfo{volume}{66}},
  \bibinfo{pages}{519} (\bibinfo{year}{2011}).

\bibitem[{\citenamefont{Liang et~al.}(2015)\citenamefont{Liang, Meng, and
  Zhou}}]{Liang2015PRe570.1}
\bibinfo{author}{\bibfnamefont{H.~Z.} \bibnamefont{Liang}},
  \bibinfo{author}{\bibfnamefont{J.}~\bibnamefont{Meng}}, \bibnamefont{and}
  \bibinfo{author}{\bibfnamefont{S.-G.} \bibnamefont{Zhou}},
  \bibinfo{journal}{Phys. Rep} \textbf{\bibinfo{volume}{570}},
  \bibinfo{pages}{1} (\bibinfo{year}{2015}).

\bibitem[{\citenamefont{Meng et~al.}(2006{\natexlab{b}})\citenamefont{Meng,
  Peng, Zhang, and Zhou}}]{Meng2006PRC73.037303}
\bibinfo{author}{\bibfnamefont{J.}~\bibnamefont{Meng}},
  \bibinfo{author}{\bibfnamefont{J.}~\bibnamefont{Peng}},
  \bibinfo{author}{\bibfnamefont{S.}~\bibnamefont{Zhang}}, \bibnamefont{and}
  \bibinfo{author}{\bibfnamefont{S.-G.} \bibnamefont{Zhou}},
  \bibinfo{journal}{Phys. Rev.} \textbf{\bibinfo{volume}{C 73}},
  \bibinfo{pages}{037303} (\bibinfo{year}{2006}{\natexlab{b}}).

\bibitem[{\citenamefont{Mayer}(1949)}]{Mayer1949Phys.Rev.75.1969}
\bibinfo{author}{\bibfnamefont{M.~G.} \bibnamefont{Mayer}},
  \bibinfo{journal}{Phys.Rev} \textbf{\bibinfo{volume}{75}},
  \bibinfo{pages}{1969} (\bibinfo{year}{1949}).

\bibitem[{\citenamefont{Haxel et~al.}(1949)\citenamefont{Haxel, Jensen, and
  Suess}}]{Haxel1949Phys.Rev.75.1766}
\bibinfo{author}{\bibfnamefont{O.}~\bibnamefont{Haxel}},
  \bibinfo{author}{\bibfnamefont{J.~H.~D.} \bibnamefont{Jensen}},
  \bibnamefont{and} \bibinfo{author}{\bibfnamefont{H.~E.} \bibnamefont{Suess}},
  \bibinfo{journal}{Phys.Rev} \textbf{\bibinfo{volume}{75}},
  \bibinfo{pages}{1766} (\bibinfo{year}{1949}).

\bibitem[{\citenamefont{Meng et~al.}(1999)\citenamefont{Meng, Sugawara-Tanabe,
  Yamaji, and Arima}}]{Meng1999PRC59.154}
\bibinfo{author}{\bibfnamefont{J.}~\bibnamefont{Meng}},
  \bibinfo{author}{\bibfnamefont{K.}~\bibnamefont{Sugawara-Tanabe}},
  \bibinfo{author}{\bibfnamefont{S.}~\bibnamefont{Yamaji}}, \bibnamefont{and}
  \bibinfo{author}{\bibfnamefont{A.}~\bibnamefont{Arima}},
  \bibinfo{journal}{Phys. Rev. C} \textbf{\bibinfo{volume}{59}},
  \bibinfo{pages}{154} (\bibinfo{year}{1999}).

\bibitem[{\citenamefont{Hecht and Adler}(1969)}]{Hecht1969NPA137.129}
\bibinfo{author}{\bibfnamefont{K.~T.} \bibnamefont{Hecht}} \bibnamefont{and}
  \bibinfo{author}{\bibfnamefont{A.}~\bibnamefont{Adler}},
  \bibinfo{journal}{Nucl. Phys. A} \textbf{\bibinfo{volume}{137}},
  \bibinfo{pages}{129} (\bibinfo{year}{1969}).

\bibitem[{\citenamefont{Arima et~al.}(1969)\citenamefont{Arima, Harvey, and
  Shimizu}}]{Arima1969PLB30.517}
\bibinfo{author}{\bibfnamefont{A.}~\bibnamefont{Arima}},
  \bibinfo{author}{\bibfnamefont{M.}~\bibnamefont{Harvey}}, \bibnamefont{and}
  \bibinfo{author}{\bibfnamefont{K.}~\bibnamefont{Shimizu}},
  \bibinfo{journal}{Phys. Lett. B} \textbf{\bibinfo{volume}{30}},
  \bibinfo{pages}{517} (\bibinfo{year}{1969}).

\bibitem[{\citenamefont{Ginocchio}(1997)}]{Ginocchio1997PRL78.436}
\bibinfo{author}{\bibfnamefont{J.~N.} \bibnamefont{Ginocchio}},
  \bibinfo{journal}{Phys. Rev. Lett.} \textbf{\bibinfo{volume}{78}},
  \bibinfo{pages}{436} (\bibinfo{year}{1997}).

\bibitem[{\citenamefont{Ginocchio and Madland}(1998)}]{Ginocchio1998PRC57.1167}
\bibinfo{author}{\bibfnamefont{J.~N.} \bibnamefont{Ginocchio}}
  \bibnamefont{and} \bibinfo{author}{\bibfnamefont{D.~G.}
  \bibnamefont{Madland}}, \bibinfo{journal}{Phys. Rev. C}
  \textbf{\bibinfo{volume}{57}}, \bibinfo{pages}{1167} (\bibinfo{year}{1998}).

\bibitem[{\citenamefont{Ginocchio and Leviatan}(1998)}]{Ginocchio1998PLB425.1}
\bibinfo{author}{\bibfnamefont{J.~N.} \bibnamefont{Ginocchio}}
  \bibnamefont{and} \bibinfo{author}{\bibfnamefont{A.}~\bibnamefont{Leviatan}},
  \bibinfo{journal}{Phys. Lett. B} \textbf{\bibinfo{volume}{425}},
  \bibinfo{pages}{1} (\bibinfo{year}{1998}).

\bibitem[{\citenamefont{Ginocchio}(1999)}]{Ginocchio1999PhyRep315.231}
\bibinfo{author}{\bibfnamefont{J.~N.} \bibnamefont{Ginocchio}},
  \bibinfo{journal}{Phys. Rep} \textbf{\bibinfo{volume}{315}},
  \bibinfo{pages}{231} (\bibinfo{year}{1999}).

\bibitem[{\citenamefont{Meng et~al.}(1998)\citenamefont{Meng, Sugawara-Tanabe,
  Yamaji, Ring, and Arima}}]{Meng1998PRC58.R628}
\bibinfo{author}{\bibfnamefont{J.}~\bibnamefont{Meng}},
  \bibinfo{author}{\bibfnamefont{K.}~\bibnamefont{Sugawara-Tanabe}},
  \bibinfo{author}{\bibfnamefont{S.}~\bibnamefont{Yamaji}},
  \bibinfo{author}{\bibfnamefont{P.}~\bibnamefont{Ring}}, \bibnamefont{and}
  \bibinfo{author}{\bibfnamefont{A.}~\bibnamefont{Arima}},
  \bibinfo{journal}{Phys. Rev. C} \textbf{\bibinfo{volume}{58}},
  \bibinfo{pages}{R628} (\bibinfo{year}{1998}).

\bibitem[{\citenamefont{Chen et~al.}(2003)\citenamefont{Chen, Lv, Meng, Zhang,
  and Zhou}}]{Chen2003CPL20.358}
\bibinfo{author}{\bibfnamefont{T.~S.} \bibnamefont{Chen}},
  \bibinfo{author}{\bibfnamefont{H.}~\bibnamefont{Lv}},
  \bibinfo{author}{\bibfnamefont{J.}~\bibnamefont{Meng}},
  \bibinfo{author}{\bibfnamefont{S.-Q.} \bibnamefont{Zhang}}, \bibnamefont{and}
  \bibinfo{author}{\bibfnamefont{S.-G.} \bibnamefont{Zhou}},
  \bibinfo{journal}{Chin. Phys. Lett.} \textbf{\bibinfo{volume}{20}},
  \bibinfo{pages}{358} (\bibinfo{year}{2003}).

\bibitem[{\citenamefont{Zhou et~al.}(2003)\citenamefont{Zhou, Meng, and
  Ring}}]{Zhou2003PRL91.262501}
\bibinfo{author}{\bibfnamefont{S.-G.} \bibnamefont{Zhou}},
  \bibinfo{author}{\bibfnamefont{J.}~\bibnamefont{Meng}}, \bibnamefont{and}
  \bibinfo{author}{\bibfnamefont{P.}~\bibnamefont{Ring}},
  \bibinfo{journal}{Phys. Rev. Lett.} \textbf{\bibinfo{volume}{91}},
  \bibinfo{pages}{262501} (\bibinfo{year}{2003}).

\bibitem[{\citenamefont{Boguta and Bodmer}(1977)}]{Boguta1977NPA292.413}
\bibinfo{author}{\bibfnamefont{J.}~\bibnamefont{Boguta}} \bibnamefont{and}
  \bibinfo{author}{\bibfnamefont{A.~R.} \bibnamefont{Bodmer}},
  \bibinfo{journal}{Nucl. Phys. A} \textbf{\bibinfo{volume}{292}},
  \bibinfo{pages}{413} (\bibinfo{year}{1977}).

\bibitem[{\citenamefont{Sugahara and Toki}(1994)}]{Sugahara1994NPA579.557TM1}
\bibinfo{author}{\bibfnamefont{Y.}~\bibnamefont{Sugahara}} \bibnamefont{and}
  \bibinfo{author}{\bibfnamefont{H.}~\bibnamefont{Toki}},
  \bibinfo{journal}{Nucl. Phys. A} \textbf{\bibinfo{volume}{579}},
  \bibinfo{pages}{557} (\bibinfo{year}{1994}).

\bibitem[{\citenamefont{Long et~al.}(2004)\citenamefont{Long, Meng, Van~Giai,
  and Zhou}}]{Long2004PRC69.034319}
\bibinfo{author}{\bibfnamefont{W.~H.} \bibnamefont{Long}},
  \bibinfo{author}{\bibfnamefont{J.}~\bibnamefont{Meng}},
  \bibinfo{author}{\bibfnamefont{N.}~\bibnamefont{Van~Giai}}, \bibnamefont{and}
  \bibinfo{author}{\bibfnamefont{S.~G.} \bibnamefont{Zhou}},
  \bibinfo{journal}{Phys. Rev. C} \textbf{\bibinfo{volume}{69}},
  \bibinfo{pages}{034319} (\bibinfo{year}{2004}).

\bibitem[{\citenamefont{Brockmann and Toki}(1992)}]{Brockmann1992PRL68.3408}
\bibinfo{author}{\bibfnamefont{R.}~\bibnamefont{Brockmann}} \bibnamefont{and}
  \bibinfo{author}{\bibfnamefont{H.}~\bibnamefont{Toki}},
  \bibinfo{journal}{Phys. Rev. Lett.} \textbf{\bibinfo{volume}{68}},
  \bibinfo{pages}{3408} (\bibinfo{year}{1992}).

\bibitem[{\citenamefont{Lenske and Fuchs}(1995)}]{Lenske1995PLB345.355}
\bibinfo{author}{\bibfnamefont{H.}~\bibnamefont{Lenske}} \bibnamefont{and}
  \bibinfo{author}{\bibfnamefont{C.}~\bibnamefont{Fuchs}},
  \bibinfo{journal}{Phys. Lett. B} \textbf{\bibinfo{volume}{345}},
  \bibinfo{pages}{355} (\bibinfo{year}{1995}).

\bibitem[{\citenamefont{Fuchs et~al.}(1995)\citenamefont{Fuchs, Lenske, and
  Wolter}}]{Fuchs1995PRC52.3043}
\bibinfo{author}{\bibfnamefont{C.}~\bibnamefont{Fuchs}},
  \bibinfo{author}{\bibfnamefont{H.}~\bibnamefont{Lenske}}, \bibnamefont{and}
  \bibinfo{author}{\bibfnamefont{H.~H.} \bibnamefont{Wolter}},
  \bibinfo{journal}{Phys. Rev. C} \textbf{\bibinfo{volume}{52}},
  \bibinfo{pages}{3043} (\bibinfo{year}{1995}).

\bibitem[{\citenamefont{Typel and Wolter}(1999)}]{Typel1999NPA656.331TW99}
\bibinfo{author}{\bibfnamefont{S.}~\bibnamefont{Typel}} \bibnamefont{and}
  \bibinfo{author}{\bibfnamefont{H.~H.} \bibnamefont{Wolter}},
  \bibinfo{journal}{Nucl. Phys. A} \textbf{\bibinfo{volume}{656}},
  \bibinfo{pages}{331} (\bibinfo{year}{1999}).

\bibitem[{\citenamefont{Bouyssy et~al.}(1987)\citenamefont{Bouyssy, Mathiot,
  Van~Giai, and Marcos}}]{Bouyssy1987PRC36.380}
\bibinfo{author}{\bibfnamefont{A.}~\bibnamefont{Bouyssy}},
  \bibinfo{author}{\bibfnamefont{J.-F.} \bibnamefont{Mathiot}},
  \bibinfo{author}{\bibfnamefont{N.}~\bibnamefont{Van~Giai}}, \bibnamefont{and}
  \bibinfo{author}{\bibfnamefont{S.}~\bibnamefont{Marcos}},
  \bibinfo{journal}{Phys. Rev. C} \textbf{\bibinfo{volume}{36}},
  \bibinfo{pages}{380} (\bibinfo{year}{1987}).

\bibitem[{\citenamefont{Bernardos et~al.}(1993)\citenamefont{Bernardos,
  Fomenko, Giai, Quelle, Marcos, Niembro, and
  Savushkin}}]{Bernardos1993PRC48.2665}
\bibinfo{author}{\bibfnamefont{P.}~\bibnamefont{Bernardos}},
  \bibinfo{author}{\bibfnamefont{V.~N.} \bibnamefont{Fomenko}},
  \bibinfo{author}{\bibfnamefont{N.~V.} \bibnamefont{Giai}},
  \bibinfo{author}{\bibfnamefont{M.~L.} \bibnamefont{Quelle}},
  \bibinfo{author}{\bibfnamefont{S.}~\bibnamefont{Marcos}},
  \bibinfo{author}{\bibfnamefont{R.}~\bibnamefont{Niembro}}, \bibnamefont{and}
  \bibinfo{author}{\bibfnamefont{L.~N.} \bibnamefont{Savushkin}},
  \bibinfo{journal}{Phys. Rev. C} \textbf{\bibinfo{volume}{48}},
  \bibinfo{pages}{2665} (\bibinfo{year}{1993}).

\bibitem[{\citenamefont{Marcos et~al.}(2004)\citenamefont{Marcos, Savushkin,
  Fomenko, L$\acute{\text{o}}$pez-Quelle, and Niembro}}]{Marcos2004JPG30.703}
\bibinfo{author}{\bibfnamefont{S.}~\bibnamefont{Marcos}},
  \bibinfo{author}{\bibfnamefont{L.~N.} \bibnamefont{Savushkin}},
  \bibinfo{author}{\bibfnamefont{V.~N.} \bibnamefont{Fomenko}},
  \bibinfo{author}{\bibfnamefont{M.}~\bibnamefont{L$\acute{\text{o}}$pez-Quelle}},
  \bibnamefont{and} \bibinfo{author}{\bibfnamefont{R.}~\bibnamefont{Niembro}},
  \bibinfo{journal}{J. Phys. G: Nucl. Part. Phys}
  \textbf{\bibinfo{volume}{30}}, \bibinfo{pages}{703} (\bibinfo{year}{2004}).

\bibitem[{\citenamefont{Long et~al.}(2006{\natexlab{a}})\citenamefont{Long,
  Giai, and Meng}}]{Long2006PLB640.150}
\bibinfo{author}{\bibfnamefont{W.~H.} \bibnamefont{Long}},
  \bibinfo{author}{\bibfnamefont{N.~V.} \bibnamefont{Giai}}, \bibnamefont{and}
  \bibinfo{author}{\bibfnamefont{J.}~\bibnamefont{Meng}},
  \bibinfo{journal}{Phys. Lett. B} \textbf{\bibinfo{volume}{640}},
  \bibinfo{pages}{150} (\bibinfo{year}{2006}{\natexlab{a}}).

\bibitem[{\citenamefont{Long et~al.}(2007)\citenamefont{Long, Sagawa, Giai, and
  Meng}}]{Long2007PRC76.034314}
\bibinfo{author}{\bibfnamefont{W.~H.} \bibnamefont{Long}},
  \bibinfo{author}{\bibfnamefont{H.}~\bibnamefont{Sagawa}},
  \bibinfo{author}{\bibfnamefont{N.~V.} \bibnamefont{Giai}}, \bibnamefont{and}
  \bibinfo{author}{\bibfnamefont{J.}~\bibnamefont{Meng}},
  \bibinfo{journal}{Phys. Rev. C} \textbf{\bibinfo{volume}{76}},
  \bibinfo{pages}{034314} (\bibinfo{year}{2007}).

\bibitem[{\citenamefont{Long et~al.}(2010{\natexlab{a}})\citenamefont{Long,
  Ring, Van~Giai, and Meng}}]{Long2010PRC81.024308}
\bibinfo{author}{\bibfnamefont{W.~H.} \bibnamefont{Long}},
  \bibinfo{author}{\bibfnamefont{P.}~\bibnamefont{Ring}},
  \bibinfo{author}{\bibfnamefont{N.}~\bibnamefont{Van~Giai}}, \bibnamefont{and}
  \bibinfo{author}{\bibfnamefont{J.}~\bibnamefont{Meng}},
  \bibinfo{journal}{Phys. Rev. C} \textbf{\bibinfo{volume}{81}},
  \bibinfo{pages}{024308} (\bibinfo{year}{2010}{\natexlab{a}}).

\bibitem[{\citenamefont{Long et~al.}(2008)\citenamefont{Long, Sagawa, Meng, and
  Giai}}]{Long2008EPL82.12001}
\bibinfo{author}{\bibfnamefont{W.~H.} \bibnamefont{Long}},
  \bibinfo{author}{\bibfnamefont{H.}~\bibnamefont{Sagawa}},
  \bibinfo{author}{\bibfnamefont{J.}~\bibnamefont{Meng}}, \bibnamefont{and}
  \bibinfo{author}{\bibfnamefont{N.~V.} \bibnamefont{Giai}},
  \bibinfo{journal}{Europhys Lett.} \textbf{\bibinfo{volume}{82}},
  \bibinfo{pages}{12001} (\bibinfo{year}{2008}).

\bibitem[{\citenamefont{Long et~al.}(2009)\citenamefont{Long, Nakatsukasa,
  Sagawa, Meng, Nakada, and Zhang}}]{Long2009PLB680.428}
\bibinfo{author}{\bibfnamefont{W.~H.} \bibnamefont{Long}},
  \bibinfo{author}{\bibfnamefont{T.}~\bibnamefont{Nakatsukasa}},
  \bibinfo{author}{\bibfnamefont{H.}~\bibnamefont{Sagawa}},
  \bibinfo{author}{\bibfnamefont{J.}~\bibnamefont{Meng}},
  \bibinfo{author}{\bibfnamefont{H.}~\bibnamefont{Nakada}}, \bibnamefont{and}
  \bibinfo{author}{\bibfnamefont{Y.}~\bibnamefont{Zhang}},
  \bibinfo{journal}{Phys. Lett. B} \textbf{\bibinfo{volume}{680}},
  \bibinfo{pages}{428} (\bibinfo{year}{2009}).

\bibitem[{\citenamefont{Li et~al.}(2016)\citenamefont{Li, Margueron, Long, and
  Giai}}]{Li2016PLB753.97}
\bibinfo{author}{\bibfnamefont{J.~J.} \bibnamefont{Li}},
  \bibinfo{author}{\bibfnamefont{J.}~\bibnamefont{Margueron}},
  \bibinfo{author}{\bibfnamefont{W.~H.} \bibnamefont{Long}}, \bibnamefont{and}
  \bibinfo{author}{\bibfnamefont{N.~V.} \bibnamefont{Giai}},
  \bibinfo{journal}{Phys. Lett. B} \textbf{\bibinfo{volume}{753}},
  \bibinfo{pages}{97} (\bibinfo{year}{2016}).

\bibitem[{\citenamefont{Wang et~al.}(2013)\citenamefont{Wang, Dong, and
  Long}}]{Wang2013PRC87.047301}
\bibinfo{author}{\bibfnamefont{L.~J.} \bibnamefont{Wang}},
  \bibinfo{author}{\bibfnamefont{J.~M.} \bibnamefont{Dong}}, \bibnamefont{and}
  \bibinfo{author}{\bibfnamefont{W.~H.} \bibnamefont{Long}},
  \bibinfo{journal}{Phys. Rev. C} \textbf{\bibinfo{volume}{87}},
  \bibinfo{pages}{047301} (\bibinfo{year}{2013}).

\bibitem[{\citenamefont{Jiang et~al.}(2015{\natexlab{a}})\citenamefont{Jiang,
  Yang, Dong, and Long}}]{Jiang2015PRC91.025802}
\bibinfo{author}{\bibfnamefont{L.~J.} \bibnamefont{Jiang}},
  \bibinfo{author}{\bibfnamefont{S.}~\bibnamefont{Yang}},
  \bibinfo{author}{\bibfnamefont{J.~M.} \bibnamefont{Dong}}, \bibnamefont{and}
  \bibinfo{author}{\bibfnamefont{W.~H.} \bibnamefont{Long}},
  \bibinfo{journal}{Phys. Rev. C} \textbf{\bibinfo{volume}{91}},
  \bibinfo{pages}{025802} (\bibinfo{year}{2015}{\natexlab{a}}).

\bibitem[{\citenamefont{Jiang et~al.}(2015{\natexlab{b}})\citenamefont{Jiang,
  Yang, Sun, Long, and Gu}}]{Jiang2015PRC91.034326}
\bibinfo{author}{\bibfnamefont{L.~J.} \bibnamefont{Jiang}},
  \bibinfo{author}{\bibfnamefont{S.}~\bibnamefont{Yang}},
  \bibinfo{author}{\bibfnamefont{B.~Y.} \bibnamefont{Sun}},
  \bibinfo{author}{\bibfnamefont{W.~H.} \bibnamefont{Long}}, \bibnamefont{and}
  \bibinfo{author}{\bibfnamefont{H.~Q.} \bibnamefont{Gu}},
  \bibinfo{journal}{Phys. Rev.} \textbf{\bibinfo{volume}{C 91}},
  \bibinfo{pages}{034326} (\bibinfo{year}{2015}{\natexlab{b}}).

\bibitem[{\citenamefont{Zong and Sun}(2018)}]{Zong2018CPC42.024101}
\bibinfo{author}{\bibfnamefont{Y.-Y.} \bibnamefont{Zong}} \bibnamefont{and}
  \bibinfo{author}{\bibfnamefont{B.-Y.} \bibnamefont{Sun}},
  \bibinfo{journal}{Chin. Phys. C} \textbf{\bibinfo{volume}{42}},
  \bibinfo{pages}{024101} (\bibinfo{year}{2018}).

\bibitem[{\citenamefont{Liang et~al.}(2008)\citenamefont{Liang, Van~Giai, and
  Meng}}]{Liang2008PRL101.122502}
\bibinfo{author}{\bibfnamefont{H.~Z.} \bibnamefont{Liang}},
  \bibinfo{author}{\bibfnamefont{N.}~\bibnamefont{Van~Giai}}, \bibnamefont{and}
  \bibinfo{author}{\bibfnamefont{J.}~\bibnamefont{Meng}},
  \bibinfo{journal}{Phys. Rev. Lett.} \textbf{\bibinfo{volume}{101}},
  \bibinfo{pages}{122502} (\bibinfo{year}{2008}).

\bibitem[{\citenamefont{Liang et~al.}(2009)\citenamefont{Liang, Van~Giai, and
  Meng}}]{Liang2009PRC79.064316}
\bibinfo{author}{\bibfnamefont{H.~Z.} \bibnamefont{Liang}},
  \bibinfo{author}{\bibfnamefont{N.}~\bibnamefont{Van~Giai}}, \bibnamefont{and}
  \bibinfo{author}{\bibfnamefont{J.}~\bibnamefont{Meng}},
  \bibinfo{journal}{Phys. Rev.} \textbf{\bibinfo{volume}{C 79}},
  \bibinfo{pages}{064316} (\bibinfo{year}{2009}).

\bibitem[{\citenamefont{Liang et~al.}(2012)\citenamefont{Liang, Zhao, and
  Meng}}]{Liang2012PRC85.064302}
\bibinfo{author}{\bibfnamefont{H.~Z.} \bibnamefont{Liang}},
  \bibinfo{author}{\bibfnamefont{P.~W.} \bibnamefont{Zhao}}, \bibnamefont{and}
  \bibinfo{author}{\bibfnamefont{J.}~\bibnamefont{Meng}},
  \bibinfo{journal}{Phys. Rev.} \textbf{\bibinfo{volume}{C 85}},
  \bibinfo{pages}{064302} (\bibinfo{year}{2012}).

\bibitem[{\citenamefont{Niu et~al.}(2013)\citenamefont{Niu, Niu, Liang, Long,
  Nik$\check{\text{s}}$i$\acute{\text{c}}$, Vretenar, and
  Meng}}]{Niu2013PLB723.172}
\bibinfo{author}{\bibfnamefont{Z.~M.} \bibnamefont{Niu}},
  \bibinfo{author}{\bibfnamefont{Y.~F.} \bibnamefont{Niu}},
  \bibinfo{author}{\bibfnamefont{H.~Z.} \bibnamefont{Liang}},
  \bibinfo{author}{\bibfnamefont{W.~H.} \bibnamefont{Long}},
  \bibinfo{author}{\bibfnamefont{T.}~\bibnamefont{Nik$\check{\text{s}}$i$\acute{\text{c}}$}},
  \bibinfo{author}{\bibfnamefont{D.}~\bibnamefont{Vretenar}}, \bibnamefont{and}
  \bibinfo{author}{\bibfnamefont{J.}~\bibnamefont{Meng}},
  \bibinfo{journal}{Phys. Lett. B} \textbf{\bibinfo{volume}{723}},
  \bibinfo{pages}{172} (\bibinfo{year}{2013}).

\bibitem[{\citenamefont{Niu et~al.}(2017)\citenamefont{Niu, Niu, Liang, Long,
  and Meng}}]{Niu2017PRC95.044301}
\bibinfo{author}{\bibfnamefont{Z.~M.} \bibnamefont{Niu}},
  \bibinfo{author}{\bibfnamefont{Y.~F.} \bibnamefont{Niu}},
  \bibinfo{author}{\bibfnamefont{H.~Z.} \bibnamefont{Liang}},
  \bibinfo{author}{\bibfnamefont{W.~H.} \bibnamefont{Long}}, \bibnamefont{and}
  \bibinfo{author}{\bibfnamefont{J.}~\bibnamefont{Meng}},
  \bibinfo{journal}{Phys. Rev. C} \textbf{\bibinfo{volume}{95}},
  \bibinfo{pages}{044301} (\bibinfo{year}{2017}).

\bibitem[{\citenamefont{Sun et~al.}(2008)\citenamefont{Sun, Long, Meng, and
  Lombardo}}]{Sun2008PRC78.065805}
\bibinfo{author}{\bibfnamefont{B.~Y.} \bibnamefont{Sun}},
  \bibinfo{author}{\bibfnamefont{W.~H.} \bibnamefont{Long}},
  \bibinfo{author}{\bibfnamefont{J.}~\bibnamefont{Meng}}, \bibnamefont{and}
  \bibinfo{author}{\bibfnamefont{U.}~\bibnamefont{Lombardo}},
  \bibinfo{journal}{Phys. Rev.} \textbf{\bibinfo{volume}{C 78}},
  \bibinfo{pages}{065805} (\bibinfo{year}{2008}).

\bibitem[{\citenamefont{Long et~al.}(2012)\citenamefont{Long, Sun, Hagino, and
  Sagawa}}]{Long2012PRC85.025806}
\bibinfo{author}{\bibfnamefont{W.~H.} \bibnamefont{Long}},
  \bibinfo{author}{\bibfnamefont{B.~Y.} \bibnamefont{Sun}},
  \bibinfo{author}{\bibfnamefont{K.}~\bibnamefont{Hagino}}, \bibnamefont{and}
  \bibinfo{author}{\bibfnamefont{H.}~\bibnamefont{Sagawa}},
  \bibinfo{journal}{Phys. Rev.} \textbf{\bibinfo{volume}{C 85}},
  \bibinfo{pages}{025806} (\bibinfo{year}{2012}).

\bibitem[{\citenamefont{Zhao et~al.}(2015)\citenamefont{Zhao, sun, and
  Long}}]{Zhao2015JPG42.095101}
\bibinfo{author}{\bibfnamefont{Q.}~\bibnamefont{Zhao}},
  \bibinfo{author}{\bibfnamefont{B.~Y.} \bibnamefont{sun}}, \bibnamefont{and}
  \bibinfo{author}{\bibfnamefont{W.~H.} \bibnamefont{Long}},
  \bibinfo{journal}{J. Phys. G: Nucl. Part. Phys.}
  \textbf{\bibinfo{volume}{42}}, \bibinfo{pages}{095101}
  (\bibinfo{year}{2015}).

\bibitem[{\citenamefont{Li et~al.}(2014)\citenamefont{Li, Long, Margueron, and
  Giai}}]{Li2014PLB732.169}
\bibinfo{author}{\bibfnamefont{J.~J.} \bibnamefont{Li}},
  \bibinfo{author}{\bibfnamefont{W.~H.} \bibnamefont{Long}},
  \bibinfo{author}{\bibfnamefont{J.}~\bibnamefont{Margueron}},
  \bibnamefont{and} \bibinfo{author}{\bibfnamefont{N.~V.} \bibnamefont{Giai}},
  \bibinfo{journal}{Phys. Lett. B} \textbf{\bibinfo{volume}{732}},
  \bibinfo{pages}{169} (\bibinfo{year}{2014}).

\bibitem[{\citenamefont{Li et~al.}(2019)\citenamefont{Li, Long, Margueron, and
  Giai}}]{Li2019PLB788.192}
\bibinfo{author}{\bibfnamefont{J.~J.} \bibnamefont{Li}},
  \bibinfo{author}{\bibfnamefont{W.~H.} \bibnamefont{Long}},
  \bibinfo{author}{\bibfnamefont{J.}~\bibnamefont{Margueron}},
  \bibnamefont{and} \bibinfo{author}{\bibfnamefont{N.~V.} \bibnamefont{Giai}},
  \bibinfo{journal}{Phys. Lett. B} \textbf{\bibinfo{volume}{788}},
  \bibinfo{pages}{192} (\bibinfo{year}{2019}).

\bibitem[{\citenamefont{Geng et~al.}(2006)\citenamefont{Geng, Meng, Hiroshi,
  Long, and Shen}}]{Geng2006CPL23.1139}
\bibinfo{author}{\bibfnamefont{L.~S.} \bibnamefont{Geng}},
  \bibinfo{author}{\bibfnamefont{J.}~\bibnamefont{Meng}},
  \bibinfo{author}{\bibfnamefont{T.}~\bibnamefont{Hiroshi}},
  \bibinfo{author}{\bibfnamefont{W.~H.} \bibnamefont{Long}}, \bibnamefont{and}
  \bibinfo{author}{\bibfnamefont{G.}~\bibnamefont{Shen}},
  \bibinfo{journal}{Chin. Phys. Lett.} \textbf{\bibinfo{volume}{23}},
  \bibinfo{pages}{1139} (\bibinfo{year}{2006}).

\bibitem[{\citenamefont{Sun et~al.}(2017)\citenamefont{Sun, Liu, Huang, Zhang,
  Wang, Liu, Ding, Gan, Ma, Yang et~al.}}]{M.D.Sun2017PLB771.303}
\bibinfo{author}{\bibfnamefont{M.~D.} \bibnamefont{Sun}},
  \bibinfo{author}{\bibfnamefont{Z.}~\bibnamefont{Liu}},
  \bibinfo{author}{\bibfnamefont{T.~H.} \bibnamefont{Huang}},
  \bibinfo{author}{\bibfnamefont{W.~Q.} \bibnamefont{Zhang}},
  \bibinfo{author}{\bibfnamefont{J.~G.} \bibnamefont{Wang}},
  \bibinfo{author}{\bibfnamefont{X.~Y.} \bibnamefont{Liu}},
  \bibinfo{author}{\bibfnamefont{B.}~\bibnamefont{Ding}},
  \bibinfo{author}{\bibfnamefont{Z.~G.} \bibnamefont{Gan}},
  \bibinfo{author}{\bibfnamefont{L.}~\bibnamefont{Ma}},
  \bibinfo{author}{\bibfnamefont{H.~B.} \bibnamefont{Yang}},
  \bibnamefont{et~al.}, \bibinfo{journal}{Phys. Lett. B}
  \textbf{\bibinfo{volume}{771}}, \bibinfo{pages}{303} (\bibinfo{year}{2017}).

\bibitem[{\citenamefont{Long et~al.}(2006{\natexlab{b}})\citenamefont{Long,
  Sagawa, Meng, and Giai}}]{Long2006PLB639.242}
\bibinfo{author}{\bibfnamefont{W.~H.} \bibnamefont{Long}},
  \bibinfo{author}{\bibfnamefont{H.}~\bibnamefont{Sagawa}},
  \bibinfo{author}{\bibfnamefont{J.}~\bibnamefont{Meng}}, \bibnamefont{and}
  \bibinfo{author}{\bibfnamefont{N.~V.} \bibnamefont{Giai}},
  \bibinfo{journal}{Phys. Lett. B} \textbf{\bibinfo{volume}{639}},
  \bibinfo{pages}{242} (\bibinfo{year}{2006}{\natexlab{b}}).

\bibitem[{\citenamefont{Long et~al.}(2010{\natexlab{b}})\citenamefont{Long,
  Ring, Meng, Van~Giai, and Bertulani}}]{Long2010PRC81.031302(R)}
\bibinfo{author}{\bibfnamefont{W.~H.} \bibnamefont{Long}},
  \bibinfo{author}{\bibfnamefont{P.}~\bibnamefont{Ring}},
  \bibinfo{author}{\bibfnamefont{J.}~\bibnamefont{Meng}},
  \bibinfo{author}{\bibfnamefont{N.}~\bibnamefont{Van~Giai}}, \bibnamefont{and}
  \bibinfo{author}{\bibfnamefont{C.~A.} \bibnamefont{Bertulani}},
  \bibinfo{journal}{Phys. Rev. C} \textbf{\bibinfo{volume}{81}},
  \bibinfo{pages}{031302(R)} (\bibinfo{year}{2010}{\natexlab{b}}).

\bibitem[{\citenamefont{Lalazissis et~al.}(2005)\citenamefont{Lalazissis,
  Nik$\check{\text{s}}$i$\acute{\text{c}}$, Vretenar, and
  Ring}}]{Lalazissis2005PRC71.024312DDME2}
\bibinfo{author}{\bibfnamefont{G.~A.} \bibnamefont{Lalazissis}},
  \bibinfo{author}{\bibfnamefont{T.}~\bibnamefont{Nik$\check{\text{s}}$i$\acute{\text{c}}$}},
  \bibinfo{author}{\bibfnamefont{D.}~\bibnamefont{Vretenar}}, \bibnamefont{and}
  \bibinfo{author}{\bibfnamefont{P.}~\bibnamefont{Ring}},
  \bibinfo{journal}{Phys. Rev. C} \textbf{\bibinfo{volume}{71}},
  \bibinfo{pages}{024312} (\bibinfo{year}{2005}).

\bibitem[{\citenamefont{Lalazissis et~al.}(2009)\citenamefont{Lalazissis,
  Karatzikos, Fossion, Arteaga, Afanasjev, and
  Ring}}]{Lalazissis2009PLB671.36NL3S}
\bibinfo{author}{\bibfnamefont{G.~A.} \bibnamefont{Lalazissis}},
  \bibinfo{author}{\bibfnamefont{S.}~\bibnamefont{Karatzikos}},
  \bibinfo{author}{\bibfnamefont{R.}~\bibnamefont{Fossion}},
  \bibinfo{author}{\bibfnamefont{D.~P.} \bibnamefont{Arteaga}},
  \bibinfo{author}{\bibfnamefont{A.}~\bibnamefont{Afanasjev}},
  \bibnamefont{and} \bibinfo{author}{\bibfnamefont{P.}~\bibnamefont{Ring}},
  \bibinfo{journal}{Phys. Lett. B} \textbf{\bibinfo{volume}{671}},
  \bibinfo{pages}{36} (\bibinfo{year}{2009}).

\bibitem[{\citenamefont{Grawe et~al.}(2007)\citenamefont{Grawe, Langanke, and
  Mart{\'i}nez-Pinedo}}]{Grawa2007Rep.Prog.Phys70.1525}
\bibinfo{author}{\bibfnamefont{H.}~\bibnamefont{Grawe}},
  \bibinfo{author}{\bibfnamefont{K.}~\bibnamefont{Langanke}}, \bibnamefont{and}
  \bibinfo{author}{\bibfnamefont{G.}~\bibnamefont{Mart{\'i}nez-Pinedo}},
  \bibinfo{journal}{Rep.Prog.Phys} \textbf{\bibinfo{volume}{70}},
  \bibinfo{pages}{1525} (\bibinfo{year}{2007}).

\bibitem[{\citenamefont{Shen et~al.}(2013)\citenamefont{Shen, Liang, Zhao,
  Zhang, and Meng}}]{Shen2013PRC88.024311}
\bibinfo{author}{\bibfnamefont{S.~H.} \bibnamefont{Shen}},
  \bibinfo{author}{\bibfnamefont{H.~Z.} \bibnamefont{Liang}},
  \bibinfo{author}{\bibfnamefont{P.~W.} \bibnamefont{Zhao}},
  \bibinfo{author}{\bibfnamefont{S.~Q.} \bibnamefont{Zhang}}, \bibnamefont{and}
  \bibinfo{author}{\bibfnamefont{J.}~\bibnamefont{Meng}},
  \bibinfo{journal}{Phys. Rev. C} \textbf{\bibinfo{volume}{88}},
  \bibinfo{pages}{024311} (\bibinfo{year}{2013}).

\bibitem[{\citenamefont{Litvinova and
  Afanasjev}(2011)}]{E.V.Litvinova2011PRC84.014305}
\bibinfo{author}{\bibfnamefont{E.~V.} \bibnamefont{Litvinova}}
  \bibnamefont{and} \bibinfo{author}{\bibfnamefont{A.~V.}
  \bibnamefont{Afanasjev}}, \bibinfo{journal}{Phys. Rev. C}
  \textbf{\bibinfo{volume}{84}}, \bibinfo{pages}{014305}
  (\bibinfo{year}{2011}).

\bibitem[{\citenamefont{Vretenar et~al.}(2002)\citenamefont{Vretenar,
  Nik$\check{\text{s}}$i$\acute{\text{c}}$, and
  Ring}}]{Vretenar2002PRC65.024321}
\bibinfo{author}{\bibfnamefont{D.}~\bibnamefont{Vretenar}},
  \bibinfo{author}{\bibfnamefont{T.}~\bibnamefont{Nik$\check{\text{s}}$i$\acute{\text{c}}$}},
  \bibnamefont{and} \bibinfo{author}{\bibfnamefont{P.}~\bibnamefont{Ring}},
  \bibinfo{journal}{Phys. Rev. C} \textbf{\bibinfo{volume}{65}},
  \bibinfo{pages}{024321} (\bibinfo{year}{2002}).

\bibitem[{\citenamefont{Roca-Maza et~al.}(2011)\citenamefont{Roca-Maza,
  Vi$\tilde{\rm n}$as, Centelles, Ring, and
  Schuck}}]{Roca-Maza2011PRC84.054309}
\bibinfo{author}{\bibfnamefont{X.}~\bibnamefont{Roca-Maza}},
  \bibinfo{author}{\bibfnamefont{X.}~\bibnamefont{Vi$\tilde{\rm n}$as}},
  \bibinfo{author}{\bibfnamefont{M.}~\bibnamefont{Centelles}},
  \bibinfo{author}{\bibfnamefont{P.}~\bibnamefont{Ring}}, \bibnamefont{and}
  \bibinfo{author}{\bibfnamefont{P.}~\bibnamefont{Schuck}},
  \bibinfo{journal}{Phys. Rev. C} \textbf{\bibinfo{volume}{84}},
  \bibinfo{pages}{054309} (\bibinfo{year}{2011}).

\end{thebibliography}
%\end{document}

%===================================================================

\end{document}